\documentclass[11pt, a4paper, twocolumn]{article}
\usepackage{amsmath}
\usepackage{graphicx}
\usepackage{amssymb}
\usepackage{epstopdf}
\usepackage{array}
\usepackage{nicefrac}
\usepackage[center]{subfigure}
\usepackage[center]{caption}
\usepackage{fullpage}
\usepackage{float}

\usepackage{bibentry}	

\begin{document}
\title{An Agent-Based Model to Explain the Emergence of Stylised Facts in Log Returns}

\author{Elena Green\footnote{corresponding author, elena@thpys.nuim.ie} \\
	Department of Theoretical Physics, National University of Ireland Maynooth, \\ Maynooth, Co. Kildare, Ireland \\
\and Daniel Heffernan\footnote{dmh@thphys.nuim.ie} 
	\\ Department of Theoretical Physics, National University of Ireland Maynooth, \\ Maynooth, Co. Kildare, Ireland \\
\and School of Theoretical Physics, Dublin Institute for Advanced Studies, Dublin 4, Ireland}
\date{}
\maketitle

\begin{abstract}

This paper outlines an agent-based model of a simple financial market in which a single asset is available for trade by three different types of traders. The model was first introduced in the PhD thesis of one of the authors, see reference~\cite{Green14phd}. The simulated log returns are examined for the presence of the stylised facts of financial data. 
The features of leptokurtosis, volatility clustering and aggregational Gaussianity are especially highlighted and studied in detail. The following ingredients are found to be essential for the production of these stylised facts: the memory of noise traders who make random trade decisions; the inclusion of technical traders that trade in line with trends in the price and the inclusion of fundamental traders who know the ``fundamental value" of the stock and trade accordingly. When these three basic types of traders are included log returns are produced with a leptokurtic distribution and volatility clustering as well as some further statistical features of empirical data. This enhances and broadens our understanding of the fundamental processes involved in the production of empirical data by the market.

\end{abstract}

\section{Introduction}
This paper outlines the construction of an Agent-Based Model (ABM). The motivation for this new model is to add to the understanding of the reasons for some of the stylised facts of financial data. It is built in the same vein as the minimal model by Alfi et al~\cite{Agent}. The goal is to reproduce the key features of financial data with an even simpler model. Although Alfi et al claim that their model is ``minimal", new models can continue to add to our understanding of the features of financial data. 

The stylised facts discussed in this paper are those of leptokurtic log returns, volatility clustering and aggregational Gaussianity. We feel that these are the most distinctive features of log returns and they are the ones we are interested in explaining. We achieve further understanding of the origins of these features by means of the ABM presented here.

%%%%%%%%%%%%%%%%%%%%%%%
\section{Building the model}
\label{build}
Where some models may investigate the effect of market microstructure on the price or log return process, the ABM presented in this paper is chiefly focused on trader behaviour and its effect on the qualitative form of the log returns. It is concerned with the stylised facts of financial data and the explanation of them from a trader-behaviour perspective. We do not attempt to create a complete realistic market microstructure. The agents in the model are built in a way that attempts to capture essential features in an extremely simplified fashion.

In this model, as in the minimal model by Alfi et al~\cite{Agent} and in the Lux and Marchesi model~\cite{LuxMarchesi00}, there is just one asset available for trade. Each trader may only buy or sell one unit of the asset at a time and trading takes place at discrete points in time. At each time step each trader might buy, sell, or stay inactive. The price is calculated from these trade decisions and all trades are executed at this price. The agents do not learn or adapt their strategies during the simulation.

The price update mechanism is multiplicative. After agents express their trade decision the excess demand $D_t$ is calculated. $D_t$ is defined as the number of buyers minus the number of sellers at time $t$; $D_t = N_{\text{buy},t} - N_{\text{sell},t}$. 
 
Following other models~\cite{Challet97, Challet01, Agent, Giardina03} the price $S_t$ at time $t$ is then generated as a function of the excess demand $D_t$ as follows
\begin{equation*}
	S_{t} =  \left( 1+ m\frac{D_t}{N} \right) S_{t-1}
\label{Price}
\end{equation*}
where $m$ is a parameter limiting the largest proportional change in the price in one iteration of the model and $N$ is the total number of traders which is fixed for the entire simulation.

The parameter $m$ measures the impact of trading on the price. $\nicefrac{D_t}{N}$ is the proportion of traders with the majority opinion ($-1 \leq \nicefrac{D_t}{N} \leq 1$). Therefore $m$ controls how much influence this majority has on the price. When $m<0$ the price moves in the opposite direction to that indicated by the demand of the agents. When $m=0$ the price is static. When $m>1$ negative prices would occur when all traders want to sell $\left(\nicefrac{D_t}{N}=-1 \right)$. Hence we restrict the parameter $m$ to $0<m<1$.

It is assumed that all traders have infinite wealth and so can always afford to buy shares. They are also given enough shares so that they always have the option to sell. It is the trade demand rather than the executed trades which influences the price process.

In the following sections we describe the trading rules of the different types of traders in the model. The model has three different types of traders operating in the market. These are noise traders who decide randomly whether to buy or sell with probability based on a certain memory of past price changes; technical traders who analyse historical prices to inform their trades, and fundamental traders who know the ``fundamental value" of the stock and trade accordingly. There are $N_1$ noise traders with a knowledge of just the most recent price change, $N_5$ with a memory of the last 5 price changes and $N_{21}$ with a memory of the last 21 price changes (think of these as day, week and month traders). There are $N_T$ technical traders and $N_F$ fundamental traders. There are a total of $N = N_1+N_5+N_{21}+N_T+N_F$ agents in the model.

Whether it is realistic to classify all traders as belonging to one of some set of predefined types may seem unlikely due to our experience of a very heterogeneous world. However, a paper by Tumminello et al.~\cite{TumminelloClusters11} goes some way to justify this classification by their finding that traders do tend to form discrete clusters which perform trades synchronised in both direction and time.

%%%%%%%%%%%%%%%%%%%
	\subsection{Noise Traders}
\label{homogeneous}
Noise traders base their trading decisions only on the current state of the market and not do not take into account any historical prices. At every iteration of the model each agent must make two decisions. Firstly, each agent decides whether or not to get involved in trading.
Allowing agents to be inactive has been found to be crucial to the presence of the stylised facts~\cite{Giardina03, Johnson00, Jefferies01, BouchaudGiardina01}. For example in the model of Alfi et al~\cite{Agent}, in an attempt to explain the self-organisation of markets into the intermittent state which produces stylised facts, agents only trade if their personal price signal is greater than a minimum threshold. 

We incorporate this concept in our model. The number of agents who are actively trading changes in response to the history of the proportional price change $R_t$, where
\begin{equation*}
R_t = \frac{S_t - S_{t-1}}{S_{t-1}}.
\label{R_t}
\end{equation*}
Agents can have various memory lengths by applying an exponential moving average (EMA) of previous proportional price changes of varying length $n$:
\begin{equation*}
R_{t,n} = w(n) R_t + \left( 1- w(n) \right) R_{t-1,n}.
\label{memoryRt}
\end{equation*}
The weight $w(n)= \frac{2}{n+1}$ where $n>0$ is the integer number of trading periods they remember. Traders who base their decisions on different memory lengths of historical prices are thought to be responsible for some of the stylised facts of empirical log returns~\cite{LeBaron06}. We include this feature by allowing the memory length $n$ to take on a variety of values for different agents.

The value $R_{t,n}$ is then used by agents to decide if they will trade or not in the current trading period according to the function $\Omega_t=\Omega(R_{t,n})$ where $\Omega_t$ is the proportion of agents that will trade after observing $R_{t,n}$,

\begin{equation}
\Omega_t = \frac{1+de^{-a(|R_{t,n}|-b)}}{1+e^{-a(|R_{t,n}|-b)}}.
\label{NumActiveTraders}
\end{equation}
$a$, $b$ and $d$ are constant parameters which will be discussed below. The number of agents with memory length $n$ that trade at time $t$ is $[ N_n \Omega(R_{t,n}) ]$ where $N_n$ is the total number of traders with memory $n$ and $[\cdot]$ denotes the nearest integer function. This new equation~\ref{NumActiveTraders} plays a fundamental role in our model. See the discussion below and in particular Section~\ref{Leptokurtic}.

\begin{figure}[h!]
	\centerline{\includegraphics[width=\linewidth]{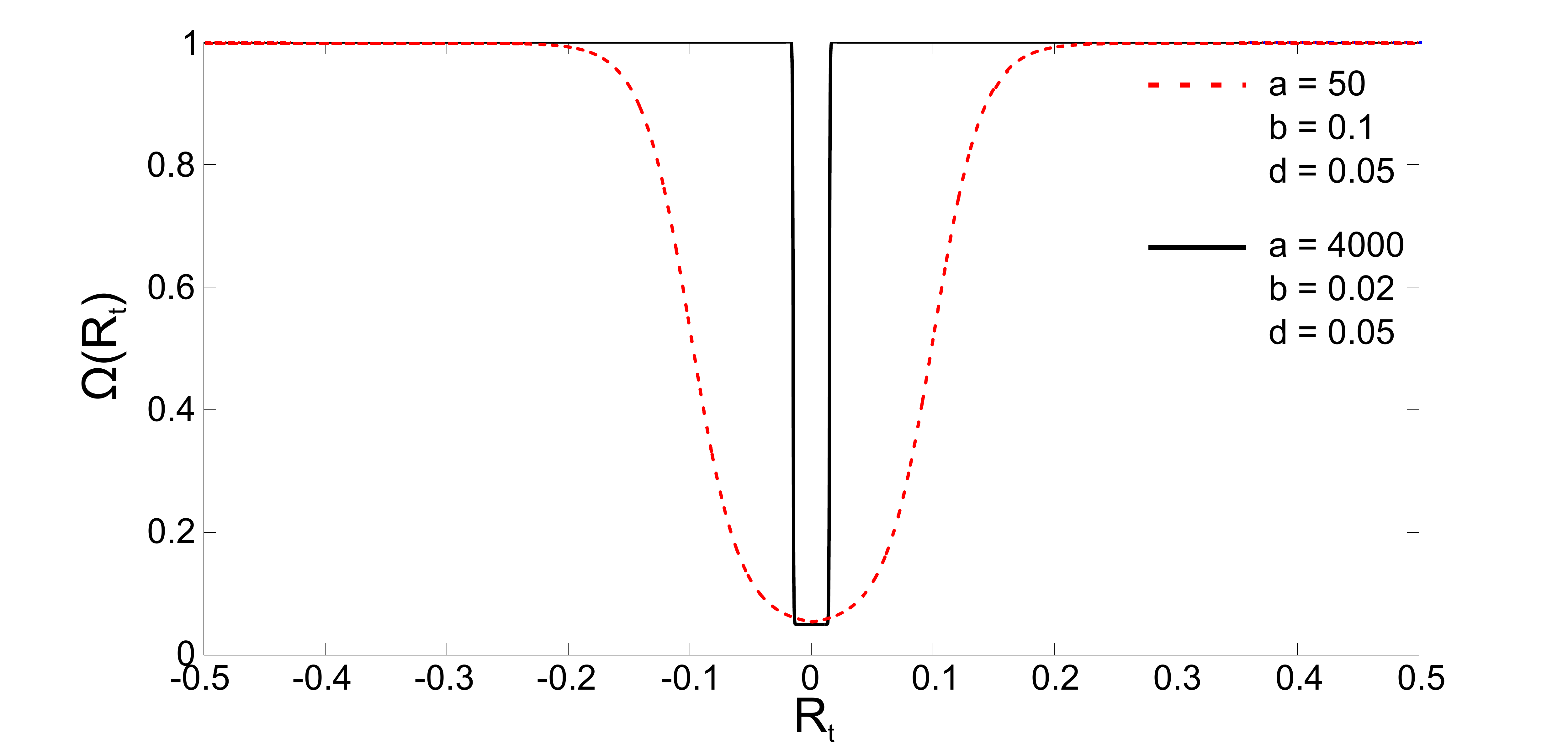}}
\caption{Graph of $\Omega$, equation~\ref{NumActiveTraders}, the proportion of active traders as a function of the latest proportional price change, shown for various parameter values as indicated on the graph.}
\label{OmegaVariations}
\end{figure}

Thus in this model it is a collective decision about the proportion of traders who trade rather than a personal decision by each trader based on a personal idiosyncratic signal. $R_{t,n}$ close to zero will lead to the minimum number of permitted active traders submitting a trade request while $R_{t,n}$ far from zero will lead to $\Omega_t = 1$ and all traders will attempt to trade. Figure~\ref{OmegaVariations} shows a graph of this function for two different parameter selections.

The number of noise agents active in the model is therefore dynamic but is bounded above by the total number of traders $N_n$ and below by
$$\left[\frac{1+de^{ab}}{1+e^{ab}} N_n \right] \approx [dN_n]$$
 for the parameter values used in this paper. The parameter $d$, $0<d<1$, controls the minimum proportion of agents who are active at any time. Since the number of active noise traders is rounded to the nearest whole number it may be $0$ if $d$ is small. 

The steepness of the function is controlled by $a$. For high values of $a$ the transition between $\Omega_t$ being at a minimum and a maximum is sharp in $R_{t,n}$, so the number of active traders is usually at one of these extremes. 
For lower $a$ there is more scope for variations in the number of active traders.

The parameter $b$ controls the width of the interval of values of $R_{t,n}$ for which the minimum number of agents are active. 
 The width of this range grows with $b$.

\begin{figure}[h!]
	\centerline{\includegraphics[width=\linewidth]{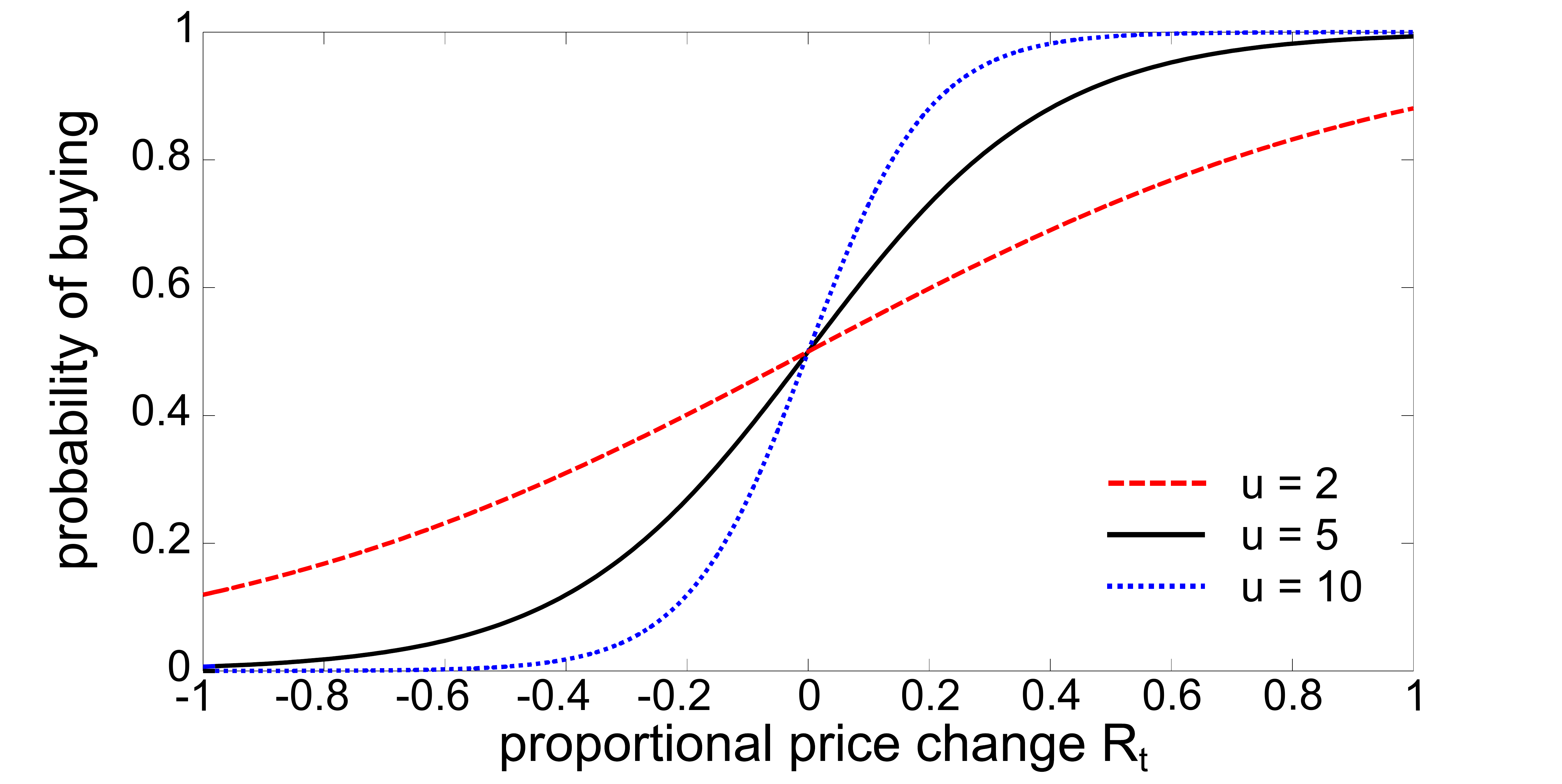}}
\caption{Graph of $P_t$, equation~\ref{buyingProb}, the probability of buying.}
\label{probBuy}
\end{figure}

The second decision that the active agents must make is whether to buy or sell. Only one share can be traded by each agent at each time step. The decision to buy or sell is made randomly according to a probability distribution $P_t$ which is based on the previous period's proportional price change $R_{t,n}$. It is the same for each noise trader with memory $n$.
\begin{equation}
	P_t = \mathbb{P}[\text{buy} | R_{t,n}] = \frac{1}{1+e^{- u R_{t,n}}}
\label{buyingProb}
\end{equation}

where $u $ controls the steepness of the function $P_t$.
$P_t$ is shown in Figure~\ref{probBuy} for various values of the parameter $u$.

The parameter $u$ can be interpreted as a ``herding strength" parameter. When $u$ is small, $P_t$ is quite flat. This means that the probability of buying will be at its extremes 0 and 1 for only very extreme $R_{t,n}$. This allows for heterogeneity among the noise traders when making their trading choices, so the herding strength is low. When $u$ is large $P_t$ is steep and so it reaches its extremes of 0 and 1 for more modest $R_{t,n}$. This means that there is more agreement in the market, or that there is strong herding among the noise traders. This causes high volatility as everyone is trading in the same direction which leads to large jumps in the price.

If $u=0$, $P_t= \nicefrac{1}{2}$ for all values of $R_{t,n}$ which is equivalent to the noise traders choosing the direction of their trades by simply tossing a coin. In order to produce dynamics which are dependent on the price moves we set $u>0$. (Having $u<0$ leads to the counterintuitive situation where noise traders are more likely to buy after observing a negative price move than a positive one. This would amount to having contrarian traders. This concept has been explored elsewhere~\cite{Grannan94, Zhong05}.)

These noise traders alone are not expected to cause volatility clustering in the log returns which result from their trades. They will instead cause clustering in the direction of trade since a positive price change increases the probability of buying which leads to another positive price change.

Following standard practice in many other agent-based models~\cite{OverviewCristelli, LuxMarchesi00, Agent}, new types of traders are introduced to the model. These are technical traders and fundamental traders. Technical traders, or chartists, analyse the price history looking for trends while fundamental traders are more concerned with the fundamental profit-generating potential of the company in which they are investing~\cite[chapter 2]{Richmond2013}.

%%%%%%%%%%%%%%%%%
	\subsection{Technical traders}
Technical traders or chartists inform their trading choices by indicators from past prices such as moving averages. They use these indicators or signals to attempt to predict future price moves. For example on a price chart, if the moving average of the price crosses over the price it shows that there has been a change in the trend. Technical traders use signals like this as a basis for their trading decisions.

The chartists in the model by Lux and Marchesi~\cite{LuxMarchesi00} are divided into optimists and pessimists. The optimists always buy and the pessimists always sell. They do not analyse historical prices at all. In the Minimal Model by Alfi et al~\cite{Agent} the chartists use a basic moving average of historical prices compared to the current price to identify trends. 

The technical traders in this model use a slightly more involved technical analysis of trends in the price in order to make their decisions as this was found to lead to more realistic results. They calculate the Moving Average Convergence Divergence (MACD). Although the aim of this model is to be as simple as possible, the rationale for using this more complicated technique lies in its realism. It also leads to richer dynamics in the results as the traders have a fuller picture of price trends than that afforded by the basic moving average. Unlike the original noise traders, the technical traders' trading decisions are completely deterministic given the price history.

The MACD technique first involves taking two EMAs of the price, $A$ and $B$, of different lengths $l_A$ and $l_B$.  Then find the difference $M_t$ between these two moving averages. This difference is called the MACD. Next calculate an EMA of the MACD, $s_t$, of length $l>0$. These steps are described by the following equations: 
\begin{align*}
A_t =&w(l_A) S_t + \left(1-w(l_A) \right)A_{t-1}  \notag\\
B_t=& w(l_B) S_t + \left(1-w(l_B) \right)B_{t-1} \notag\\
M_t =&A_t - B_t \notag\\
s_t =& w(l) M_t + \left(1-w(l)\right)s_{t-1}
\label{MACD}
\end{align*}

The weight $w$ depends on the length; $w(l) = \frac{2}{l+1}.$ A comparison between the MACD $M_t$ and its EMA $s_t$ indicates trends in the price. $M_t > s_t$ indicates that the price is on an upward trend and the technical traders will respond by buying the stock. $M_t < s_t$ shows that the price is on a downward trend and the technical traders will respond to this signal by selling. The technical traders therefore amplify the price trends they detect.

This leads to the excess demand by the technical traders:
\begin{align*}
D_t =& N_{\text{buy}} - N_{\text{sell}}\\
 =& N_T \text{sgn}(A_t - B_t - s_t)
\end{align*}
 where $N_T$ is the total number of technical traders and $\text{sgn}(x)$ is the sign function given by
\[
\text{sgn}(x) := 
\begin{cases}
1, & \quad x>0 \\
0, & \quad x=0 \\
-1, & \quad x<0.
\end{cases}
\]

There is a problem with having technical traders in the model. The amplification of trends leads the price to either grow to infinity or drop toward zero very quickly. All traders also have unlimited buying power and so extremely large prices are a common occurrence. The price can also drop to extremely small values.

Another type of trader is necessary to keep the market reasonably stable. Fundamental traders will fill this role.

%%%%%%%%%%%%%%%%%%
	\subsection{Fundamental traders}
\label{fundamental}
In order to have fundamental traders in the model, there must first of all be a defined ``fundamental value" for the traded asset. In a real trading environment, the fundamental value of a stock can be estimated as the current value of expected future dividend payments. This sort of calculation clearly cannot be performed within this model. Other models which have fundamental traders often set the fundamental value to some constant level for the duration of the simulation~\cite{LuxMarchesi00, Agent}. Allowing the fundamental value to vary or giving fundamental traders heterogeneous beliefs may allow for more interesting dynamics~\cite{AlfiII09, Ferreira05}.

In this model all the fundamental traders agree with each other on what the fundamental value $f$ is at any moment. $f$ is set to follow a random walk:
\begin{equation}
f_t = f_{t-1}\left( 1+ \mu_f + \sigma_f \epsilon_t \right)
\label{fundValue}
\end{equation}
$\mu_f$ and $\sigma_f$ are the mean and variance of $f$, and $\epsilon_t$ is a random number taken from a standard normal distribution. %The time steps are set to $\Delta t = 1$.

The fundamental traders know the fundamental value of the asset. At time $t$, they compare the price $S_t$ to $f_t$ and decide if the asset represents good value. They will buy if the price is below the fundamental value and sell if it is above. Their trading strategy pulls the price back towards the fundamental value. They have the opposite effect on prices to the technical traders and help to stabilise the market. Like the technical traders their decisions are deterministic once $f_t$ is known and $S_t$ is revealed. All of the fundamental traders trade in the same direction.

The demand of the fundamental traders at time $t$ is therefore given by
\begin{equation*}
D_t = N_F\text{sgn}(f_t - S_t)
\end{equation*}
where $N_F$ is the total number of fundamental traders in the model.

\section{Results}
In this section the log returns generated by the ABM will be tested for the stylised facts of empirical log returns. The stylised facts which have been found to be present in the synthetic log return time series include leptokurtosis, volatility clustering and aggregational Gaussianity. These features are discussed below. The number of each trader type present in the model described as Trader Sets A and B are given in Table~\ref{table_TraderSets}. The parameters used in the model are shown in Table~\ref{table_Parameters}.
{\renewcommand{\arraystretch}{1.3}
\begin{table}%[t!]
\begin{center}	
	\centerline{
	$\begin{array}{|l|c|c|c|c|c|}	% array must be in math mode. Needs to be surrounded by $
	\hline
	\text{model set-up} &  N_1 & N_5 & N_{21} & N_T & N_F\\ \hline
	\text{Trader Set A } & 4 & 4 & 8 & 2 & 2 \\ 
	\text{Trader Set B } &  0 & 0 & 16 & 2 & 2 \\ 
 \hline
	\end{array}$
	}
\caption{The number of the different types of traders in the ABM for the results presented below. $N_1$, $N_5$ and $N_{21}$ are noise traders with memories of 1, 5, and 21 times steps respectively.}
\label{table_TraderSets}
\end{center}
\end{table}

\begin{table*}[t!]
\begin{center}
	\centerline{
	$\begin{array}{|c||c||c|c|c||c|c|c||c|c|}	% array must be in math mode. Needs to be surrounded by $
	\hline
	 S & P & \multicolumn{3}{c||}{\Omega} & \multicolumn{3}{c||}{\text{MACD}} & \multicolumn{2}{c|}{f}  \\ \hline
	m & u & a & b  & d & l_A & l_B & l & \mu_f & \sigma_f \\ \hline
	0.4 & 5 & 4000 & 0.02 & 0.05 & 12 & 26 & 9 & 3 \cdot 10^{-4} & 0.025\\ 
	 \hline
	\end{array}$
	}
\caption{The parameters used for the ABM}
\label{table_Parameters}
\end{center}
\end{table*}

\subsection{Leptokurtic Distribution of Log Returns}
\label{Leptokurtic}
The log returns $Z$ produced by the model have been found to be consistently leptokurtically distributed independent to the presence of technical and fundamental traders. See examples in Figure~\ref{figureReturns}. In each case the log returns produced had a leptokurtic shape and the Shapiro-Francia test\footnote{This test is suitable for leptokurtic log returns. It sorts the data into ascending order and finds the correlation between this ordered data and the expected order statistics for data from a normal distribution~\cite{Royston83}.} rejected normality at a significance level of $0.1\%$. The essential feature of the ABM which produces this stylised fact is the function $\Omega$, defined in equation~\ref{NumActiveTraders}.

\begin{figure}[h!]
	\subfigure[Trader Set A]{\includegraphics[width=0.49\linewidth]{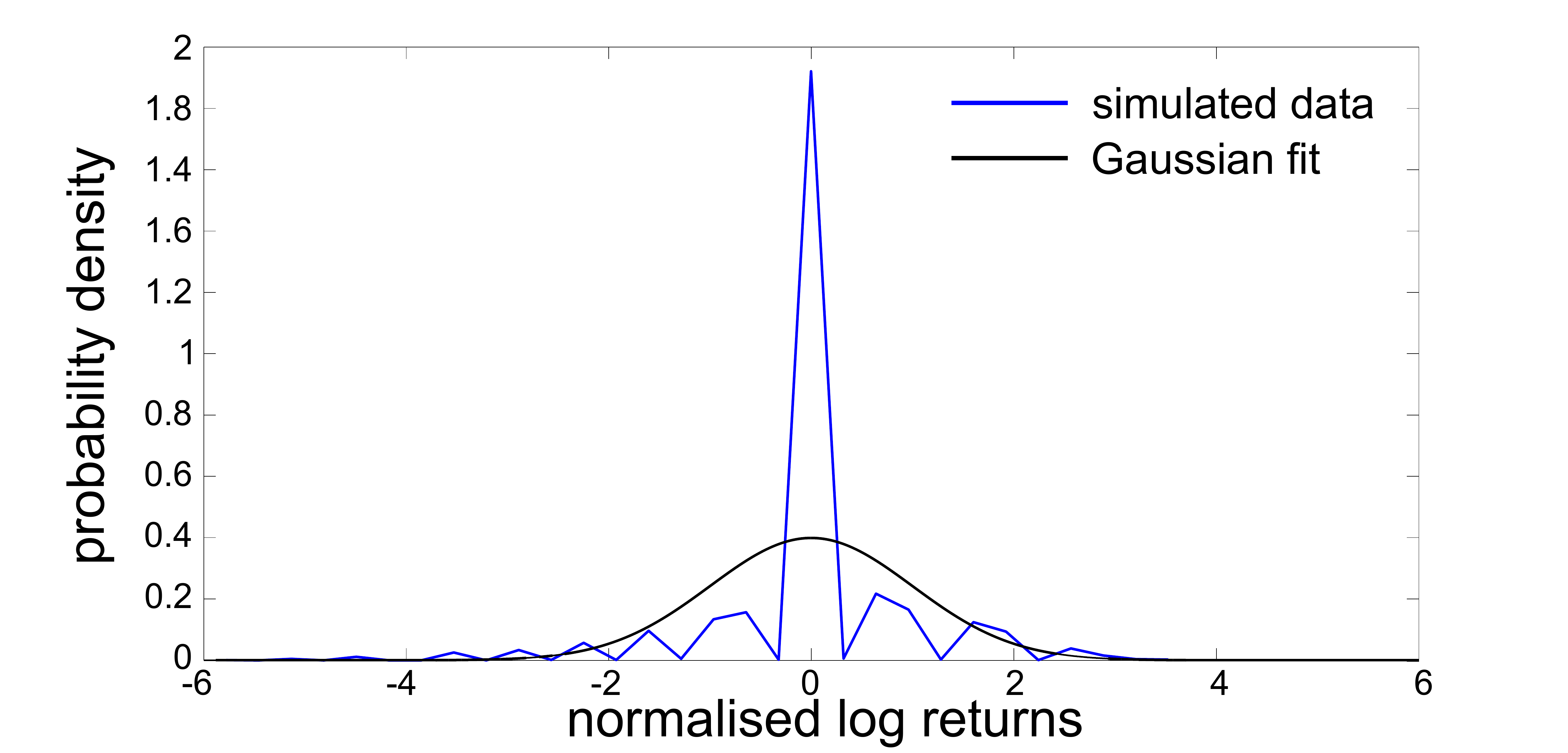}}
	\subfigure[Trader Set B]{\includegraphics[width=0.49\linewidth]{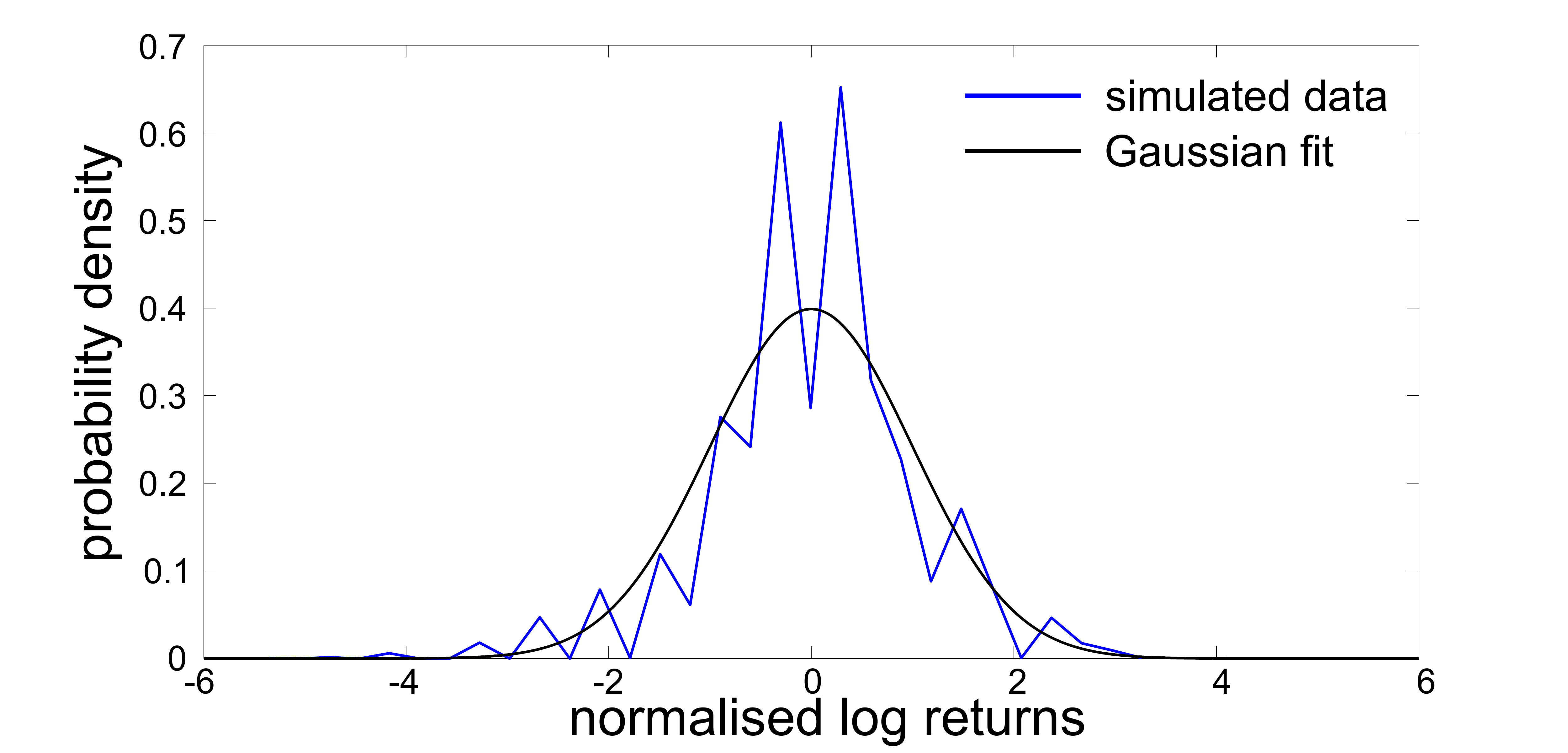}}
\caption{Examples of the distribution of normalised log returns produced by the ABM when different numbers of traders are present. A Gaussian fit is shown for comparison.}
\label{figureReturns}
\end{figure}

This function adjusts the number of active noise traders according to the previous price change. The value of $d$ is critical. If $d=1$, $\Omega=1$ and the number of noise traders active in the ABM is constant. Keeping the number of active noise traders at a constant level results in log returns which are well described by a Gaussian distribution. This is the case even when technical and fundamental traders, who are independent of $\Omega$, are also present in the ABM. 

The function $\Omega$, defined in Equation~\ref{NumActiveTraders}, mimics realistic trading patterns. In real trading if there is a large price move, perhaps as a result of some news arriving to the market, traders who are normally not very active may be motivated to review their portfolio and make some trades. This leads to more log returns close to zero when these more casual investors are not trading and extreme log returns when everybody wants to trade because they have seen a large price move.

This finding is consistent with other studies which have related the leptokutric distribution of log returns to the varying rate of trading~\cite{Clark73, Westerfield77, Gallant92, Karpoff87, Davidsson14, Zheng14}. High volatility is related to periods of high trading volume. Since within the ABM each trader can only trade one share at a time, the number of active traders $[\Omega N_N] + N_T + N_F$ is a proxy for volume.

	\subsection{Volatility Clustering}
Volatility clustering occurs in the log returns only when technical and fundamental traders are added to the market. Clusters can be identified by eye in Figure~\ref{figureReturns}.
	\begin{figure}[h]
	\subfigure[Trader Set A]{\includegraphics[width=0.49\linewidth]{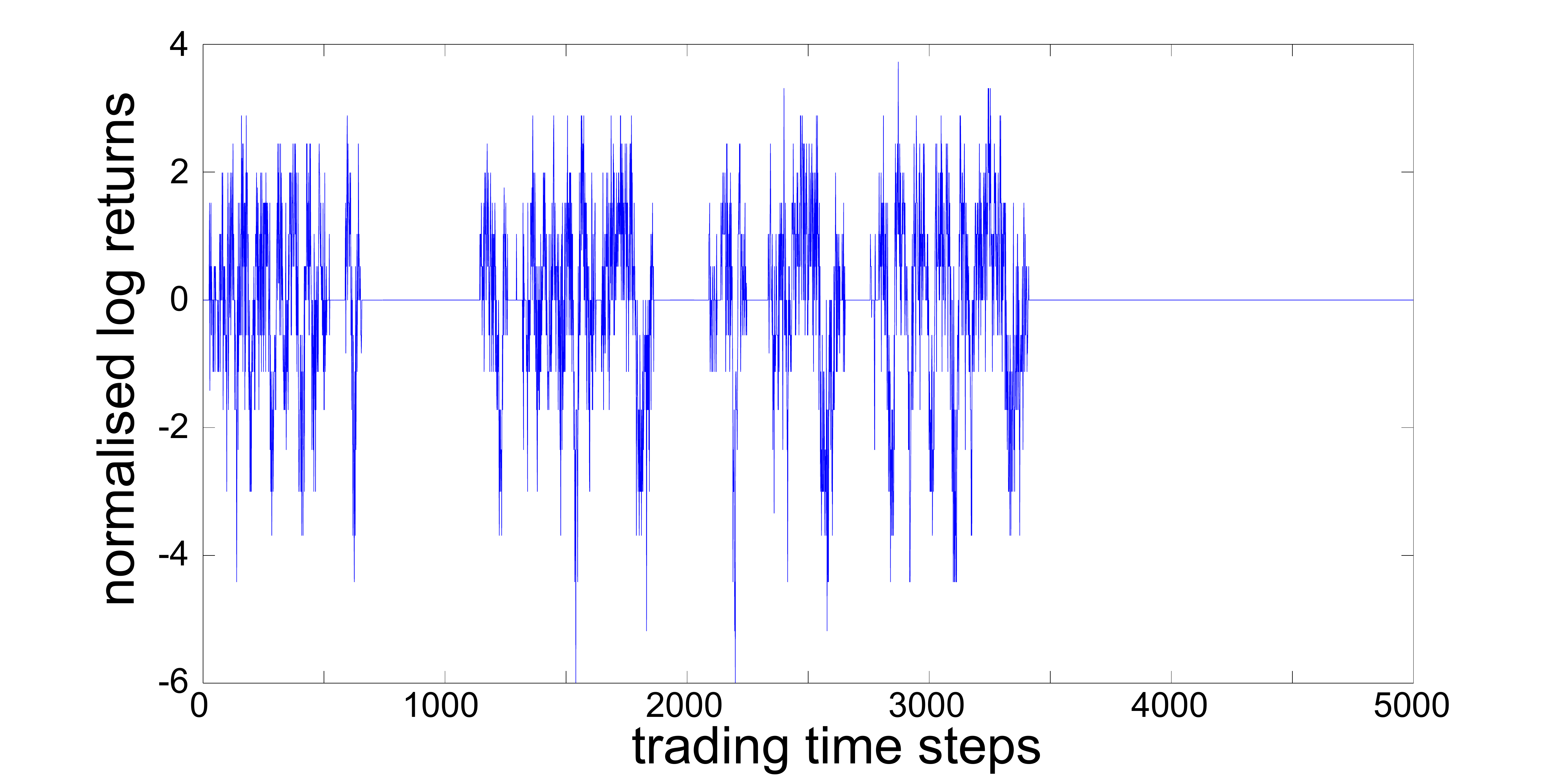}\label{modelReturnsA}}
	\subfigure[Trader Set B]{\includegraphics[width=0.49\linewidth]{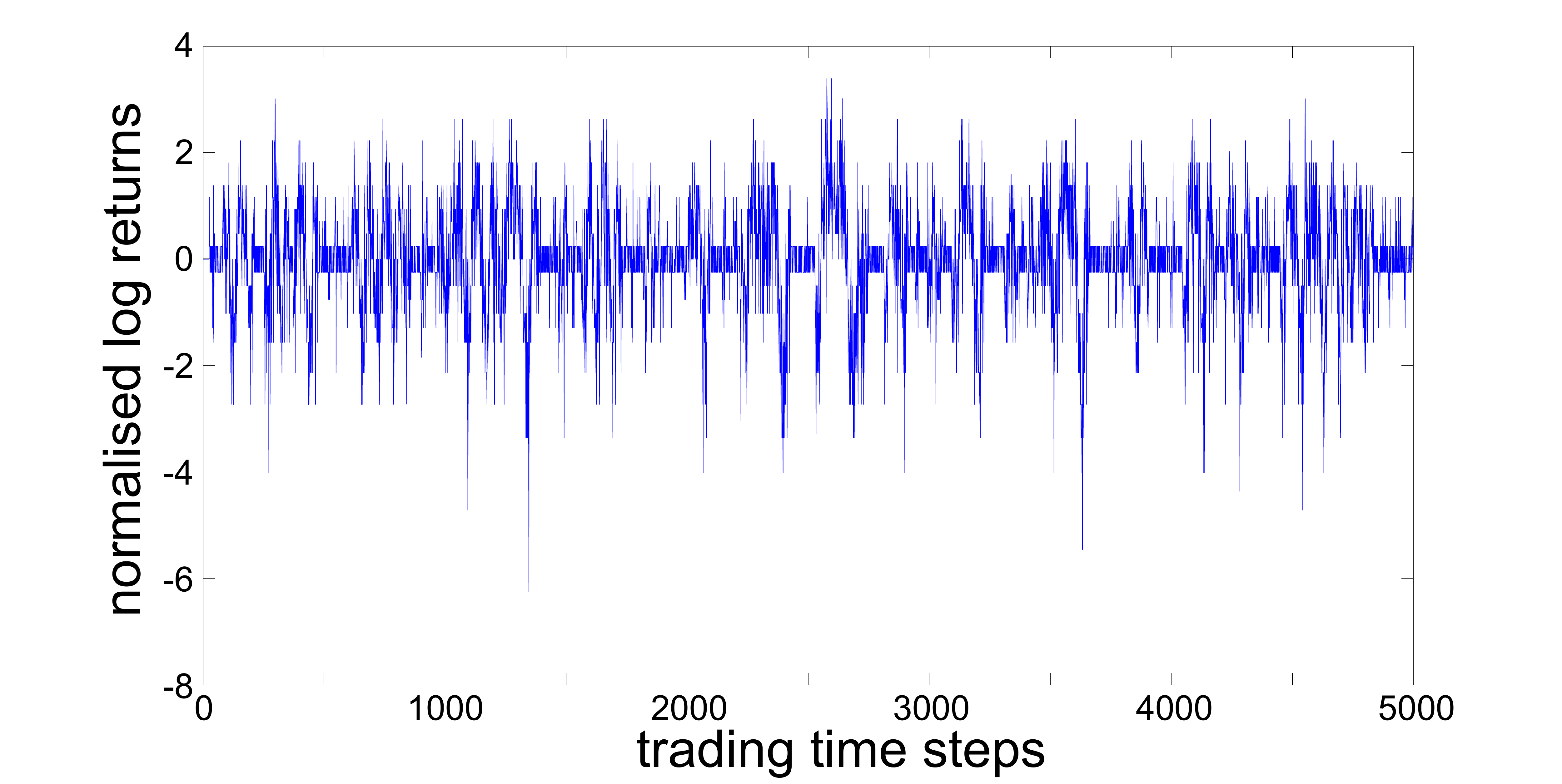}\label{modelReturnsB}}
\caption{Examples of the log returns produced by the ABM when different numbers of traders are present. They are normalised to be in units of standard deviation.}
\label{figureReturns}
\end{figure}

Examples of the autocorrelation function (ACF) are shown in Figure~\ref{figModelACF}. The magnitudes of log returns generated by Trader Sets A and B are long-term correlated. A slow decay in the ACF of absolute values of log returns is a signature of the volatility clustering that can be seen in Figure~\ref{figureReturns}.

The ACF of empirical absolute log returns decays roughly as a power law~\cite{OverviewCristelli, Giardina03}. Figure~\ref{figACFloglogDecay} shows the decay of the autocorrelation for data generated by Trader Sets A and B along with pure power laws for two different set ups of the ABM on doubly logarithmic scales. In both cases a power law provides a reasonable fit, comparable if not better than the fit to empirical data. See for example Figure 6 in ref.~\cite{OverviewCristelli}.

\begin{figure}[h!]
	\subfigure[Trader Set A]{\includegraphics[width=0.49\linewidth]{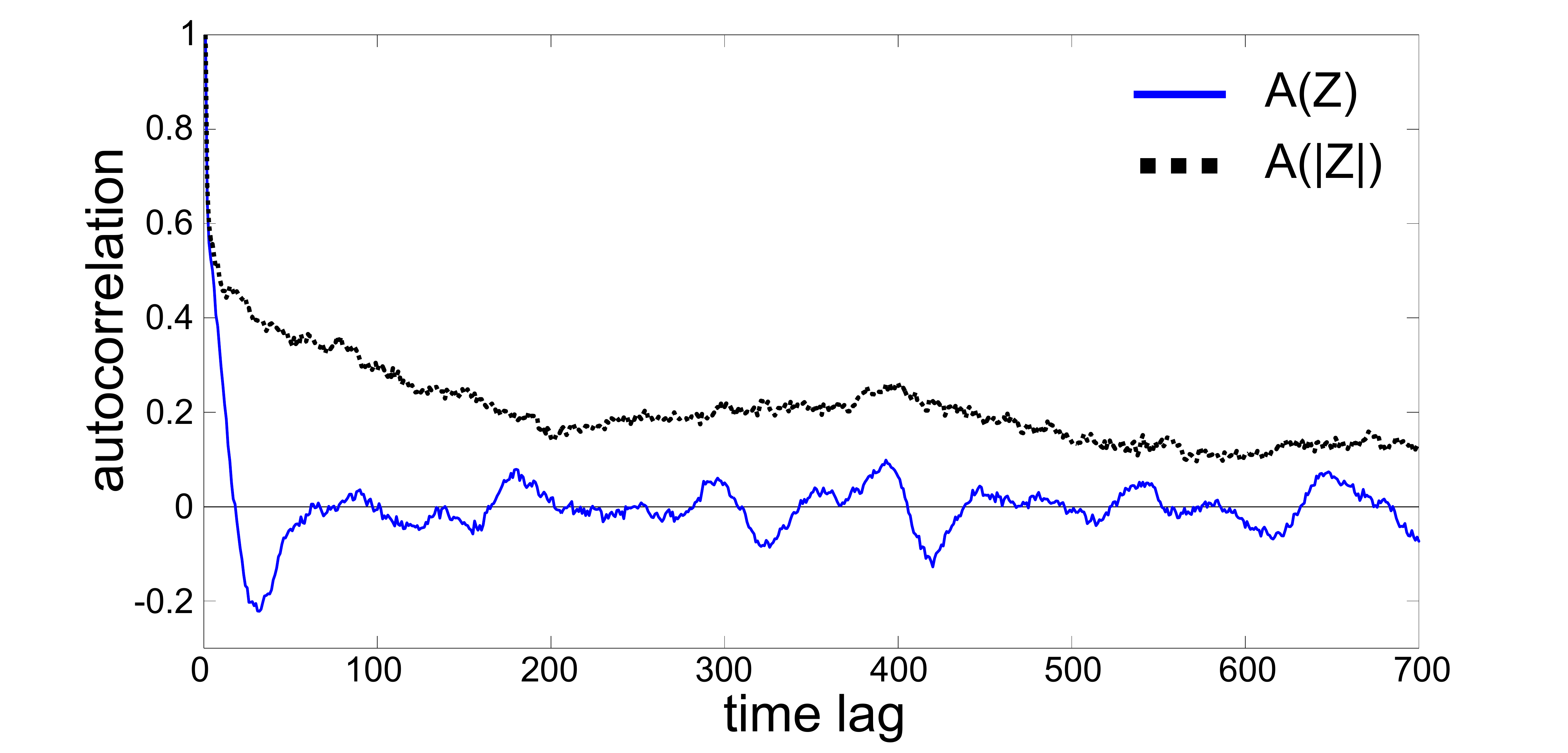}\label{modelACFA}}
	\subfigure[Trader Set B]{\includegraphics[width=0.49\linewidth]{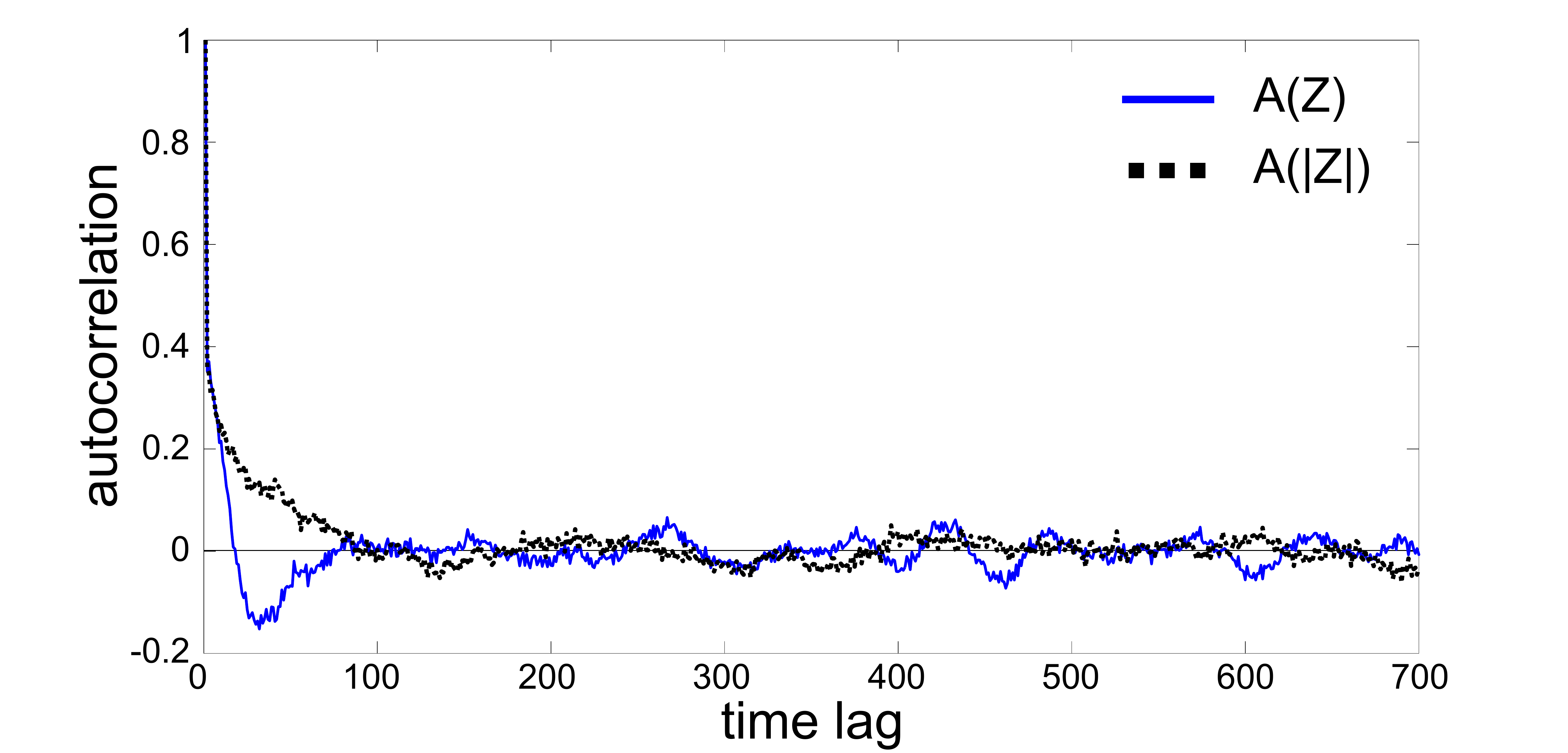}\label{modelACFB}}
\caption{Examples of the ACF of the log returns and their absolute values produced with different numbers of traders in the ABM.}
\label{figModelACF}
\end{figure}

\begin{figure}[h!]
	\centerline{\subfigure[Trader Set A]{\includegraphics[width=0.85\linewidth]{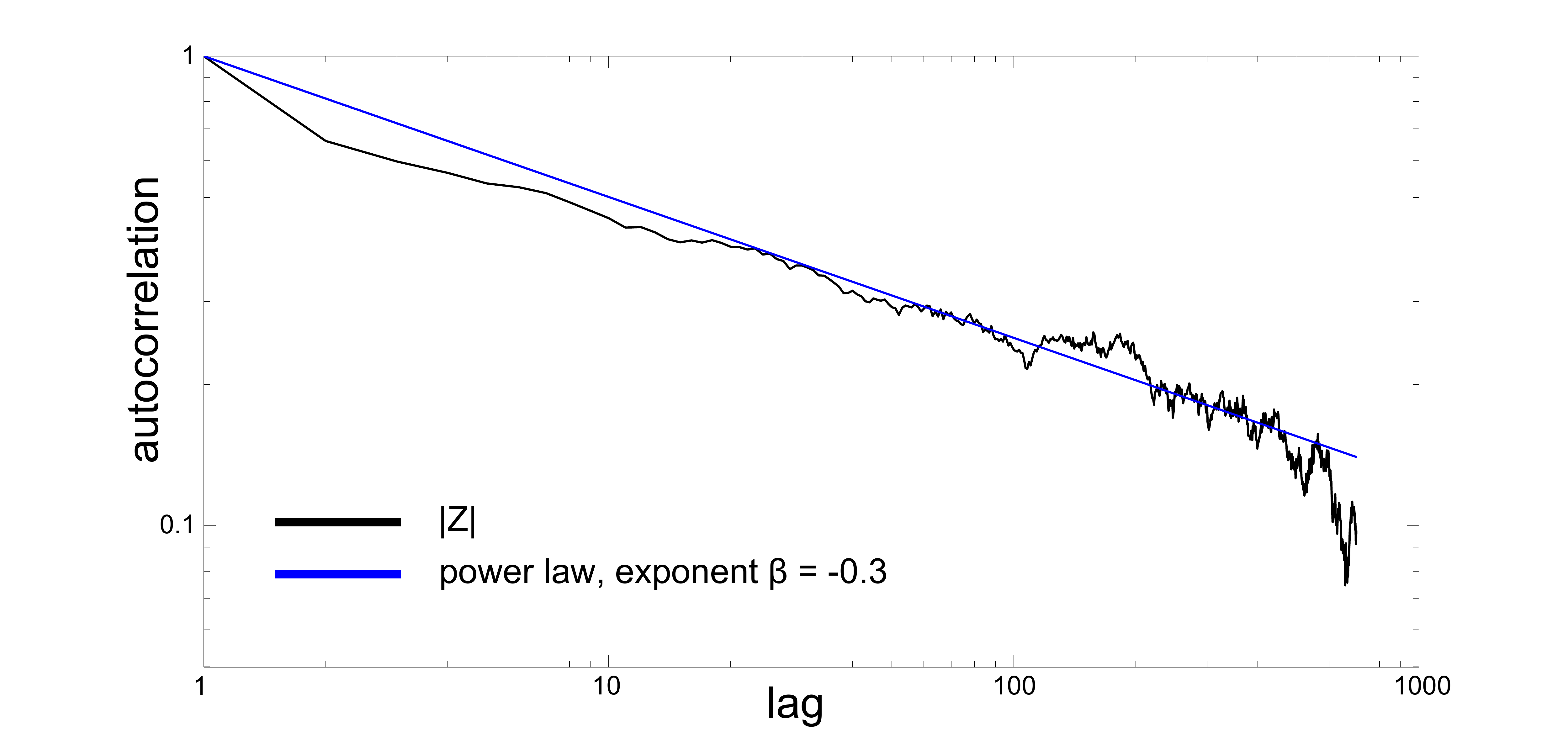}\label{figModel1ACFloglogDecay}}}
	\centerline{\subfigure[Trader Set B]{\includegraphics[width=0.85\linewidth]{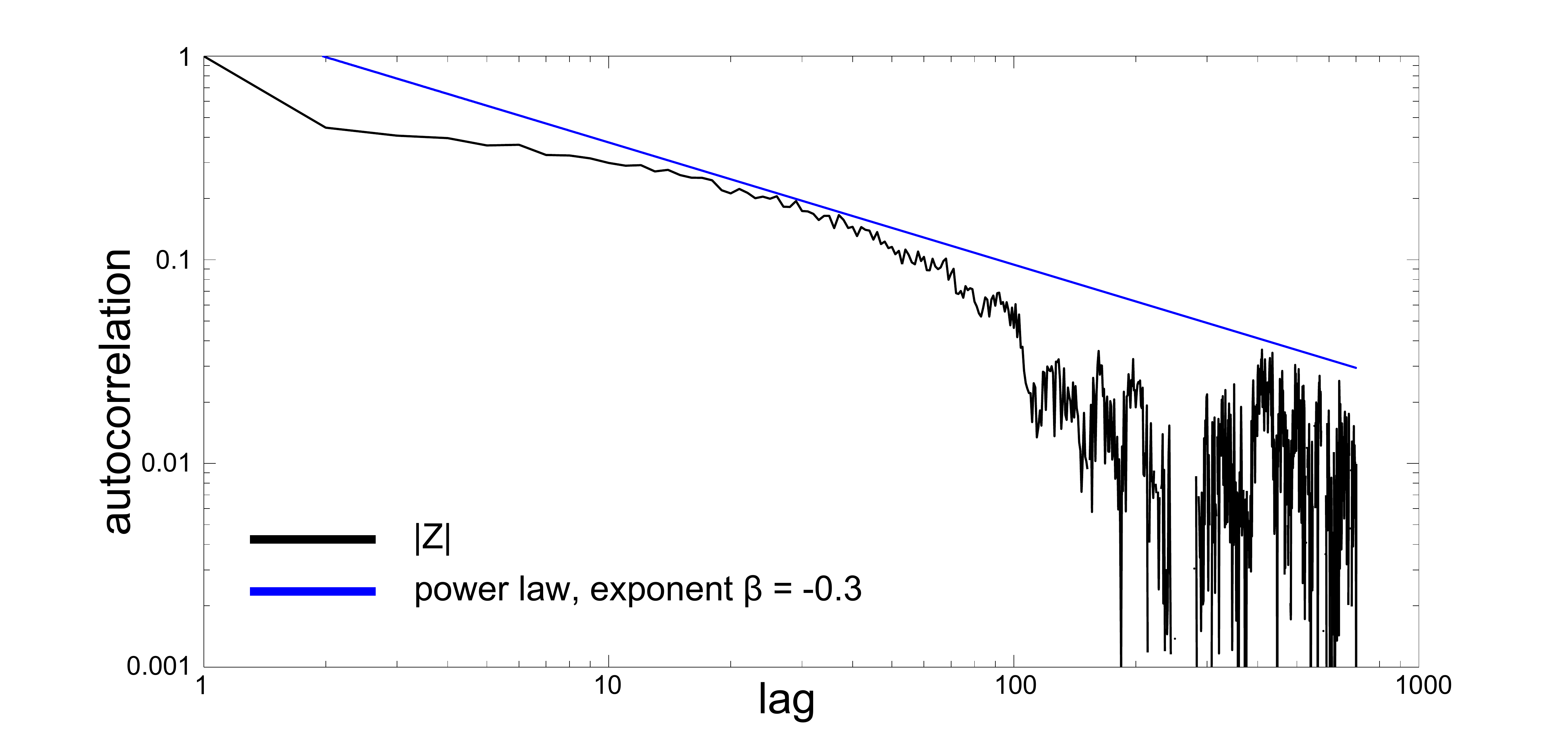}\label{figModel2ACFloglogDecay}}}
\caption{Graphs of the autocorrelation of the absolute log returns generated by the ABM on doubly logarithmic scales. Both are shown with a pure power law for comparison. The power law provides a good fit in both cases. }
\label{figACFloglogDecay}
\end{figure} 

	\subsection{Aggregational Gaussianity}
Another recognised feature of financial data is that as the time lag is increased the distribution of the log returns begins to more closely resemble a Gaussian~\cite{StylizedFacts, LeBaron06, AggGaussAntypas2013}. Specifically at long time scales, such as annual log returns, the empirical distribution is reasonably fitted by a Gaussian.

Let $ Z_{t,\Delta} = \log(S_{t+\Delta}) - \log(S_{t}). $
So far the log returns of successive prices ($\Delta = 1$) generated by the ABM have been examined. In order to look for a scale-dependent distribution log returns at different time scales $\Delta$ must be found. If $\Delta$ is allowed to increase the shape of the distribution does indeed change, as is shown in Figure~\ref{GaussianTransition}. 

\begin{figure}[h!]
	\includegraphics[width=\linewidth]{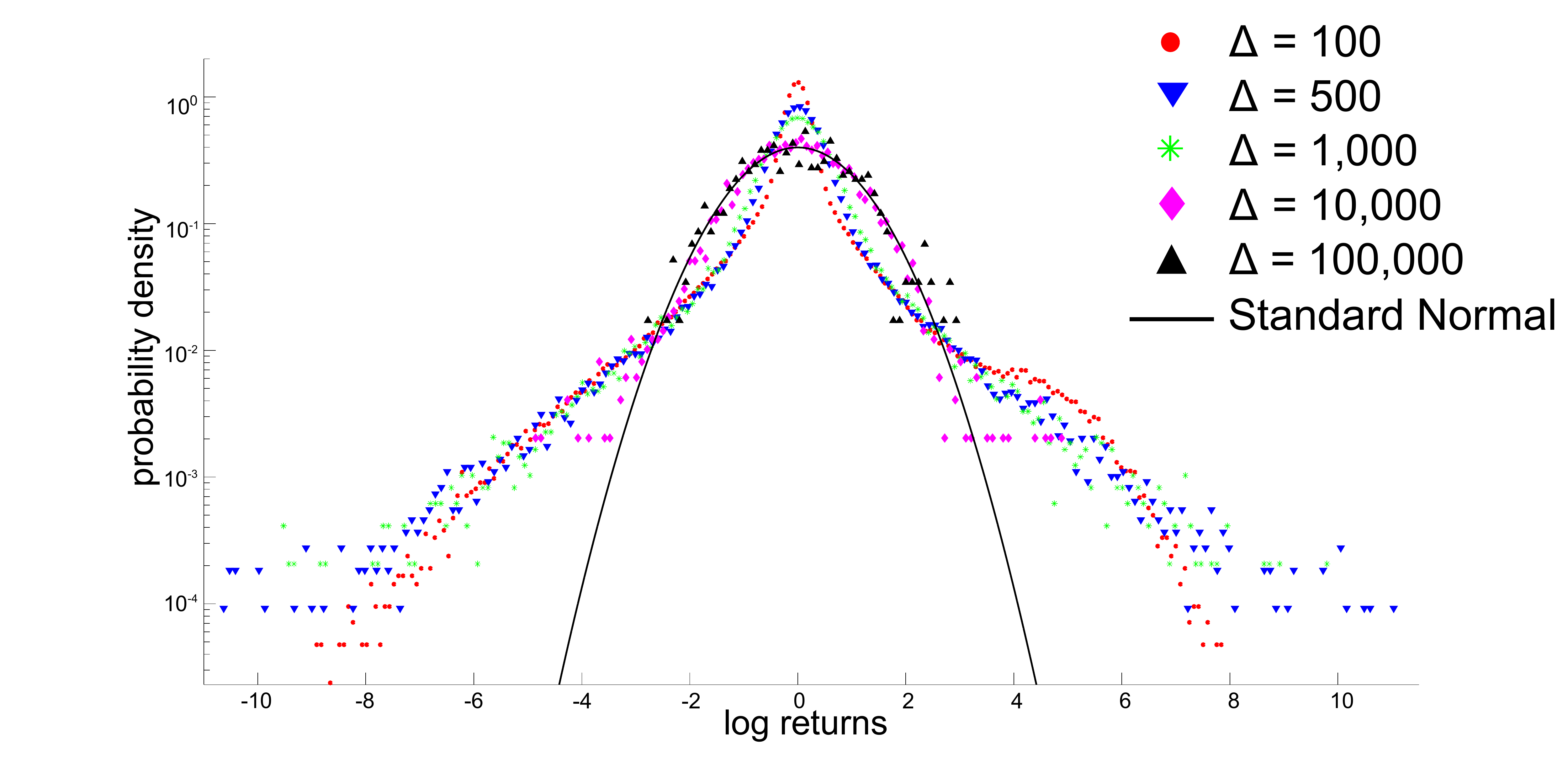}
	\caption{Graph of the distribution of normalised log returns $Z_{t,\Delta}$ calculated over different lags $\Delta$ for a long simulation ($T = 5 \cdot 10^7$) with Trader Set B and varying $f$. The solid black line shows a standard normal distribution. The vertical scale is logarithmic. }
	\label{GaussianTransition}
\end{figure}

The leptokurtic distribution begins to break down for large $\Delta$. At $\Delta = 10,000$ there is a reasonable fit to a Gaussian distribution within $3 \sigma$ of the mean, but beyond this the tails are much too fat to be explained by a Gaussian. However at $\Delta = 100,000$ all values of $Z_{t,\Delta}$ fall roughly on the Gaussian distribution. The results are shown on a semi-log scale to allow for greater visibility of the tails.

The reason for the aggregational Gaussianity lies with the fundamental value $f$. 
The log returns of $f$ have a normal distribution due to its dependence on the Gaussian random number $\epsilon_t$, see equation~\ref{fundValue}.
At large $\Delta$ large events become rare and the consistent Gaussian influence of $f$ on the fundamental traders dominates $Z_{t,\Delta}$. At large lags any short term trends instigated by technical traders are not felt and the shape of the distribution is influenced principally by the fundamental traders. 

To confirm that this is the reason for the aggregational Gaussianity, the same analysis was carried out on log returns generated by the ABM with $f$ set to a constant value for the entire simulation.
Figure~\ref{GaussianTransitionConstV} shows the result. Even at large $\Delta$ there is no agreement with a Gaussian distribution in this case. This is because there is no Gaussian influence on any traders and the log returns retain their fat tails. These results are similar to those found by Alfi et al in the analysis of their model~\cite{AlfiII09}.

\begin{figure}[h!]
	\includegraphics[width=\linewidth]{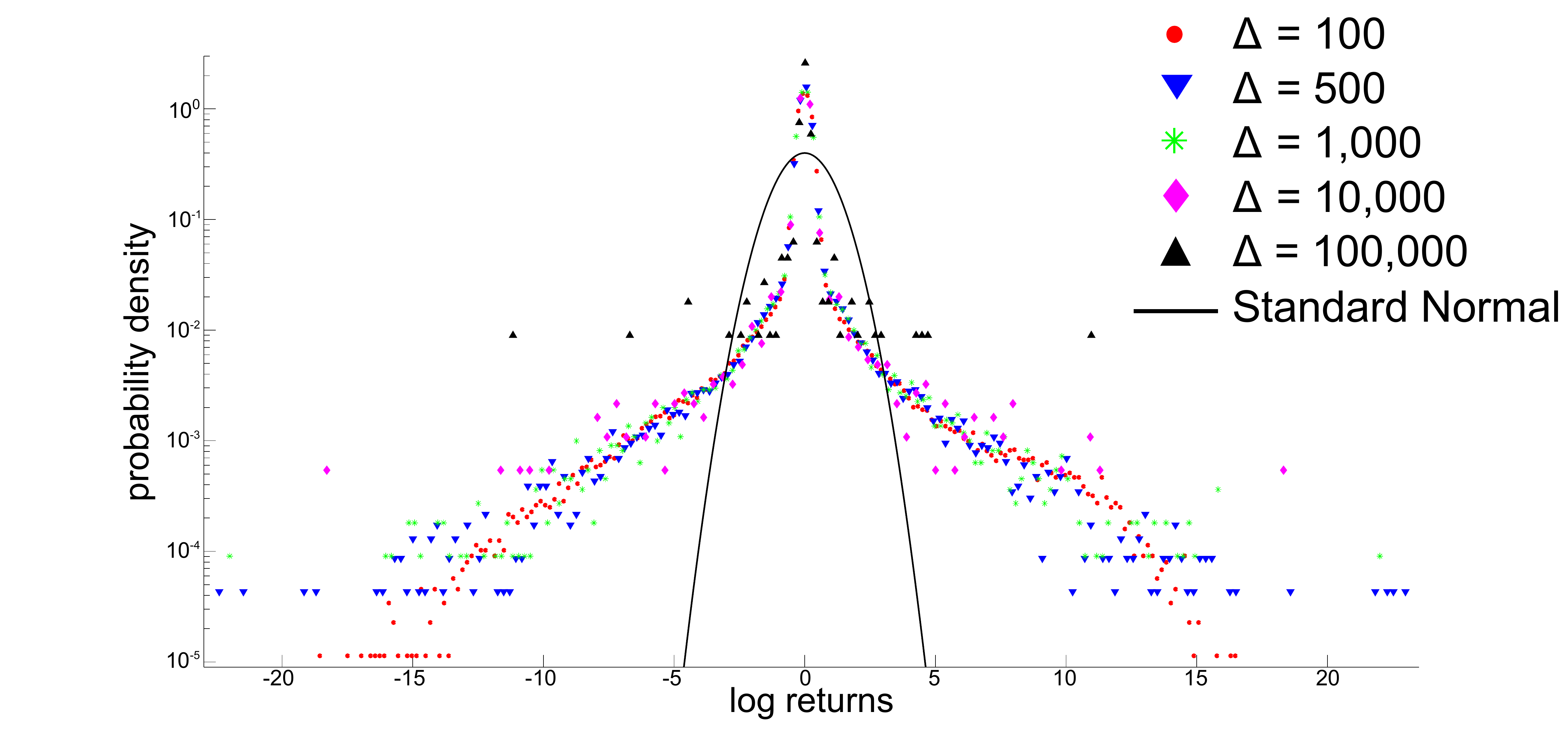}
	\caption{Graph of the distribution of normalised log returns $Z_{t,\Delta}$ calculated over different lags $\Delta$ for a long simulation ($T = 5 \cdot 10^7$) with Trader Set B and constant $f$. The solid black line shows a standard normal distribution. The vertical scale is logarithmic.}
	\label{GaussianTransitionConstV}
\end{figure}
\section{Conclusion}
\label{conc}
In this paper a new ABM has been developed. The motivation behind this new ABM was to find a very simple model which can reproduce some of the key stylised facts of empirical financial time series. This enriches the understanding of the origin of the stylised facts in empirical data. This ABM focuses on trader behaviour rather than market microstructure. As is the case with many ABMs useful results are only obtained from this model in a limited area of the parameter space~\cite{OverviewCristelli, Samanidou07}.

Leptokurtic log returns are generated by the noise traders in the ABM. It has been shown that the varying number of active traders is the source of this feature in the results. This mimics the behaviour of real traders and offers an explanation for the leptokurtosis of empirical log returns.

Volatility clusters come from  having some memory in the noise traders along with technical traders who analyse historical prices looking for patterns. Technical traders bring memory to the system as they detect trends and amplify them. 
Neither the technical traders alone nor the memory of noise traders alone is enough to produce this feature. Both of these are necessary. The presence of technical traders necessitates the presence of fundamental traders to keep the price reasonably stable. It is also the fundamental traders who trigger the bursts of high volatility. Three essential ingredients have thus been identified for this model to produce this stylised fact of financial data.  They are the memory of the noise traders, the inclusion of technical traders who trade in line with trends in the price and the inclusion of fundamental traders who trade according to the ``fundamental value" of the stock.

Transition of the distribution of the log returns to a Gaussian has also been identified as a statistical property of the log returns generated by the ABM. This is caused by the fundamental value and indicates that many real traders may also be under the influence of a GBM process.

We have shown that some of the most distinctive stylised facts of financial data have been produced by this model with just a few simple elements. This contributes to our understanding of the processes behind some of the main features of empirical data, in particular the leptokurtic distribution, volatility clustering and aggregational Gaussianity.

\paragraph{Acknowledgments}
We thank William Hanan for useful discussions on the role of Noise Traders during the early stages of this work. The research was supported in part by Science Foundation Ireland under grant number 08/SRC/FM1389.
\bibliography{C:/Users/Elena/Dropbox/work/Thesis/References} 

\end{document}